\documentclass[11pt]{article}
\usepackage[colorlinks=true]{hyperref}
\hypersetup{colorlinks=true,linkcolor=teal,citecolor=blue,urlcolor=blue}
\usepackage[miktex]{gnuplottex}
\usepackage{pgfplots}
\usepackage{graphicx}
\usepackage{amsmath}
\usepackage{amssymb}
\usepackage{latexsym}
\usepackage{color}
\usepackage{subfigure}
\usepackage{pst-all}
\usepackage{pstricks,pst-node,pst-text,pst-3d}
\usepackage{ifpdf}
\usepackage{multimedia}
\usepackage{lmodern}
\usepackage{cite}
\usepackage{dsfont}
\usepackage{bbm}
\usepackage{lipsum}
\usepackage{url}
\usepackage{textcomp}
\usepackage{bbold}
\usepackage{epstopdf}
\usepackage{epsfig}
\usepackage{mathtools}
\usepackage{nccmath}
\usepackage{float}
 \usepackage[normalem]{ulem}
\usepackage{ifpdf}
\ifpdf
\else
  \usepackage{pstricks}
\fi
  \hypersetup{}

\newcommand{\beq}{\begin{equation}}
\newcommand{\eeq}{\end{equation}}
\newcommand{\beqa}{\begin{eqnarray}}
\newcommand{\eeqa}{\end{eqnarray}}
\newcommand{\nn}{\nonumber}

\newcommand{\la}{\langle}
\newcommand{\ra}{\rangle}

\begin{document}
\thispagestyle{empty}
\begin{center}

\vspace{1.8cm}
\renewcommand{\thefootnote}{\fnsymbol{footnote}}
 {\Large {\bf Quantum dynamics of a bosonic mode and a two-level system interacting with several reservoirs}}\\

\vspace{1.5cm} {\bf Narges Cheraghpour}$^{1}$
, {\bf Fardin Kheirandish} {\footnote { {email: f.kheirandish@uok.ac.ir}}}$^{1}$\\
\vspace{0.5cm}

$^{1}${\it Department of Physics, Faculty of Science, University of Kurdistan, P.O.Box 66177-15175, Sanandaj, Iran}\\ [0.5em]

\end{center}
\baselineskip=18pt
\medskip
\vspace{3cm}
\begin{abstract}
In the framework of a novel dissipative scheme, we have investigated the quantum dynamics of an oscillating system interacting with two reservoirs with different temperatures trough different time-dependent coupling functions. The reduced density matrix, quantum optical characteristic functions, and (quasi) distribution functions like Husimi, Glauber-Sudarshan and Wigner functions on the phase space of the oscillator are obtained. The problem has been generalized to the case where the oscillator is interacting with $n$ distinctive reservoirs, and a quantum current and an effective reservoir is introduced. Finally, the quantum dynamics of a two-level system interacting with two reservoirs has been investigated, and the exact reduced density matrix is obtained.

\vspace{1cm}
\end{abstract}

\vspace{1cm}

\newpage
 \renewcommand{\thefootnote}{*}

\section{Introduction}\label{Introduction}
The Rieder-Lebowitz-Lieb model \cite{a1}, introduced in 1967, demonstrated the finite thermal conductance of a harmonic oscillator chain coupled to heat baths at different temperatures, regardless of its length. This seminal work highlighted the complexities of heat conduction in low-dimensional systems. Understanding energy transport in open systems interacting with multiple baths requires careful consideration of energy and particle exchange \cite{a2,a3,a4,a5,a6,a7}. While challenging, ideal systems offer a framework for characterizing and calculating these currents, similar to equilibrium systems \cite{a8,a10,a11}. The properties of the bath and the system-bath coupling significantly influence heat transport. Oscillator chains, serving as simplified models, have been extensively employed to investigate these effects on heat conductivity in non-equilibrium settings\cite{a12, a13,a15}.
For weakly coupled systems, the heat flux is well defined and can be directly related to the changes in the energy of the system\cite{f1,f2,f3,f4,f5}. However, this picture breaks down for strongly coupled systems driven by time-dependent processes, where changes in the system-reservoir coupling energy are significant\cite{f6}.
The quantum harmonic oscillator and two-level systems are foundational models in quantum mechanics, pivotal for understanding a wide array of physical phenomena and applications \cite{x1,x2,x3,x4}. When these systems interact with their environment, particularly through coupling with each other, they exhibit complex behaviors that are essential in fields like quantum optics, quantum information processing, and condensed matter physics. Quantum optics research has long been concerned with open quantum systems, especially those interacting with thermal baths. Thermal baths, in particular, provide a realistic model for the ubiquitous presence of noise and dissipation in experimental settings. Understanding the dynamics of open quantum systems interacting with thermal baths is crucial for various applications, including quantum information processing, quantum thermodynamics, and the development of novel quantum technologies \cite{c1,c2,c3}. At the interface of technology and fundamental physics, the study of energy transport has witnessed substantial progress in bridging classical and quantum thermodynamic frameworks. Quantum thermodynamics has emerged as a prominent area of contemporary research, offering novel insights into the behavior of quantum systems operating far from equilibrium. Recent investigations have underscored the crucial role of quantum effects in thermodynamic processes, particularly within the context of nanoscale devices and quantum information processing \cite{b1,b2,b3,b4}.

In this paper, we investigate the quantum dynamics of two widely used quantum systems, namely the harmonic oscillator and the two-level atom, interacting with multiple thermal baths with different coupling strengths and temperatures. Our approach is based on a method we introduced in \cite{d22}. We commence our investigation by delving into the dissipative harmonic oscillator, a foundational model in quantum optics, then proceed to examine a pivotal scenario: a harmonic oscillator coupled to two thermal reservoirs operating at distinct temperatures. Through rigorous mathematical techniques, we derive an exact analytical solution for the reduced density matrix of the oscillator, offering a comprehensive characterization of its quantum state. Furthermore, we leverage the power of characteristic functions to compute essential quantum distributions, including the Husimi Q-function, Glauber-Sudarshan P-function, and the Wigner function, providing deeper insights into the system's quantum behavior. Building upon these foundational results, we extend our analysis to a more intricate system encompassing an oscillator interacting with an arbitrary number of thermal reservoirs, each characterized by a unique coupling strength. Remarkably, we demonstrate that this complex system can be effectively mapped onto a simpler model involving a single thermal reservoir with an appropriately defined effective coupling strength, significantly simplifying the analytical treatment. Subsequent sections extend this investigation to a two-level system coupled to two thermal reservoirs. By deriving the exact time evolution operator for this system, we obtain the reduced density matrix, providing valuable insights into the dynamics of this paradigmatic model in quantum information science. This work aims to contribute to a deeper understanding of the impact of multiple thermal environments on the quantum dynamics of fundamental systems, with potential implications for various fields, including quantum optics, quantum thermodynamics, and quantum information processing.
\section{Basics}\label{Sec1}
\subsection{Dissipative harmonic oscillator: a novel scheme}
To model dissipation in a quantum harmonic oscillator we follow the scheme introduced in \cite{d22}. In this scheme, the environment of the main oscillator is modeled by a similar oscillator as the bath oscillator and the oscillators interact trough a time-dependent coupling function $g(t)$. Let the main oscillator be described by bosonic ladder operators $\hat{a}, \hat{a}^\dag$, and the bath oscillator be described by the ladder operators $\hat{b}, \hat{b}^\dag$. The Hamiltonian of the total system is
\begin{equation}\label{B1}
\hat{H}(t)=\hbar\omega_0\,\hat{a}^\dag\hat{a}+\hbar\omega_0\,\hat{b}^\dag\hat{b}+\hbar g(t)\,(\hat{a}^\dag\hat{b}+\hat{a}\hat{b}^\dag).
\end{equation}
Using the Bogoliubov transformations
\begin{eqnarray}\label{B2}
&& \hat{a}=\frac{1}{\sqrt{2}}(\hat{A}+\hat{B}),\nn\\
&& \hat{b}=\frac{1}{\sqrt{2}}(\hat{B}-\hat{A}),
\end{eqnarray}
the Hamiltonian becomes separable in terms of the new ladder operators
\begin{equation}\label{B3}
\hat{H}(t)=\hbar\omega_A (t)\,\hat{A}^\dag\hat{A}+\hbar\omega_B (t)\,\hat{B}^\dag\hat{B},
\end{equation}
where we defined $\omega_A (t)=\omega_0-g(t)$, $\omega_B (t)=\omega_0+g(t)$. Now from Heisenberg equations of motion and the Bogoliubiv transformations Eq.~(\ref{B2}) one easily finds
\begin{eqnarray}\label{B4}
&& \hat{a}(t)=e^{-i\omega_0 t}\cos[G(t)]\,\hat{a} (0)-i e^{-i\omega t}\sin[G(t)]\,\hat{b} (0),\nn\\
&& \hat{b}(t)=e^{-i\omega_0 t}\cos[G(t)]\,\hat{b} (0)-i e^{-i\omega t}\sin[G(t)]\,\hat{a} (0),
\end{eqnarray}
where $G(t):=\int\limits_0^t dt'\,g(t')$. For the choice $\cos[G(t)]=e^{-\frac{\gamma t}{2}}$, the results will coincide with those obtained from the Lindblad master equation \cite{d22}. Let the oscillator be initially prepared in an arbitrary state $\hat{\rho}_S (0)$ and the bath be in a thermal state with inverse temperature $\beta$
\begin{equation}\label{B5}
\hat{\rho} (0)=\hat{\rho}_S\otimes\frac{e^{-i\hbar\omega_0\hat{b}^\dag\hat{b}}}{z_b},
\end{equation}
where $z_b$ is the partition function $z_b=tr(e^{-\beta\hbar\omega_0 \hat{b}^\dag\hat{b}})$, the average number of excitations is \cite{d22}
\begin{eqnarray}\label{B6}
n(t) &=& tr \big[\hat{\rho}(0)\hat{a}^\dag (t)\hat{a} (t)\big],\nn\\
     &=& \cos^2[G(t)]\la\hat{a}^\dag(0)\hat{a}(0)\ra+\sin^2[G(t)]\la\hat{b}^\dag(0)\hat{b}(0)\ra,\nn\\
     &=& e^{-\gamma t}\,n(0)+(1-e^{-\gamma t})\,\bar{n}_b,
\end{eqnarray}
where we substituted Eqs.~(\ref{B4}) into Eq.~(\ref{B6}) and made use of $\cos^2[G(t)]=e^{-\gamma t}$, and defined
\begin{eqnarray}\label{B7}
\bar{n}_b &=& tr_b \Big[\frac{e^{-i\hbar\omega_0\hat{b}^\dag\hat{b}}}{z_b}\hat{b}^\dag(0)\hat{b}(0)\Big],\nn\\
          &=& \frac{1}{e^{\beta\hbar\omega_0}-1}.
\end{eqnarray}
For the initial state $\rho(0)=|n\ra_a\otimes|0\ra_b$, where the $a$-oscillator is initially in an excited state $|n\ra_a$ and the $b$-oscillator is initially in its ground state$|0\ra_b$, we find the corresponding energies respectively as
\begin{eqnarray}\label{energies}
  E_a (t) &=& \hbar\omega_0 (\la\hat{a}^\dag (t)\hat{a}(t)\ra+\frac{1}{2})= n e^{-\gamma t},\nn\\
  E_b(t)  &=& \hbar\omega_0 (\la\hat{b}^\dag (t)\hat{b}(t)\ra+\frac{1}{2})= n (1-e^{-\gamma t}).
\end{eqnarray}
\begin{figure}[h]
\centering
\includegraphics[scale=0.7]{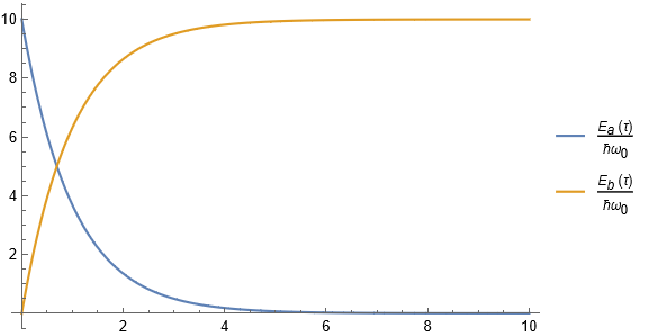}
\caption{(Color online) The scaled energies in terms of the dimensionless parameter $\tau=\gamma t$ for the initial state $|10\ra\otimes|0\ra$. The energy of the $a$-oscillator has been transferred to the $b$-oscillator (a discharge-charge process).)}\label{Fig1}
\end{figure}
In Fig.~(\ref{Fig1}), the energies of the oscillators are depicted in units of $\hbar\omega_0$ in terms of the dimensionless parameter $\tau=\gamma t$, it is seen that the $a$-oscillator has been descharged and the $b$-oscillator has been charged. Therefore, this scheme may have applications in the field of quantum batteries where an oscillator is charging by another oscillator in the presence of an external force. Note that, in the presence of an external source applied to the $a$-oscillator, the new Hamiltonian is
\begin{equation}\label{NewB1}
\hat{H}(t)=\hbar\omega_0\,\hat{a}^\dag\hat{a}+\hbar\omega_0\,\hat{b}^\dag\hat{b}+\hbar g(t)\,(\hat{a}^\dag\hat{b}+\hat{a}\hat{b}^\dag)+\hbar f_{ext} (t) (\hat{a}^\dag+\hat{a}).
\end{equation}
In this case, by making use of the same Bogoliubov transformations, one easily finds the evolution of the ladder operators from Heisenberg equations of motion as
\begin{eqnarray}
  && \hat{a}(t)=e^{-i\omega_0 t}\cos[G(t)]\,\hat{a} (0)-i e^{-i\omega_0 t}\sin[G(t)]\,\hat{b} (0)-i\int_0^t dt'\,e^{-i \omega_0 (t-t')}\cos[G(t)-G(t')]\,f_{ext} (t'),\nn\\
  && \hat{b}(t)=e^{-i\omega_0 t}\cos[G(t)]\,\hat{b} (0)-i e^{-i\omega_0 t}\sin[G(t)]\,\hat{a} (0)-\int_0^t dt'\,e^{-i \omega_0 (t-t')}\sin[G(t)-G(t')]\,f_{ext} (t'),\nn\\
\end{eqnarray}
and by making use of Eqs.~(\ref{energies}), the modified energies can be obtained.
\section{An oscillator interacting with two thermal reservoirs}\label{Sec2}
\begin{figure}[h]
\centering
\includegraphics[scale=0.15]{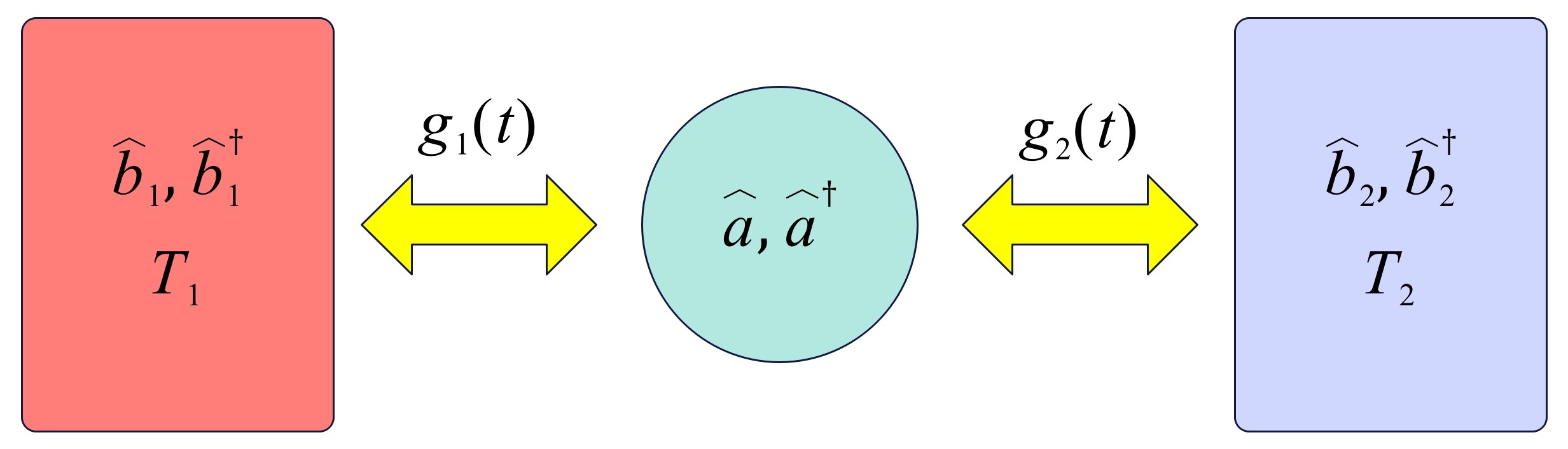}
\caption{(Color online) An oscillator interacting with two thermal reservoirs with temperatures $T_1$ and $T_2$.}\label{Fig2}
\end{figure}

Quantum heat engines utilize quantum systems, such as two-level systems or harmonic oscillators, as working substances to convert heat into work. When these systems interact with reservoirs at different temperatures, they can operate in cycles analogous to classical heat engines, but with unique quantum features. These studies have implications for the development of efficient quantum thermal machines and provide insights into the fundamental limits of energy conversion at the quantum level \cite{gelbwaser2013minimal}. In this section, we investigate the dynamics of a quantum harmonic oscillator or equivalently, a bosonic mode interacting with two thermal reservoirs with temperatures $T_1$ and $T_2$, Fig.~(\ref{Fig2}).
The Hamiltonian of the system and reservoirs can be written as \cite{d22}
\begin{equation}\label{1p}
\hat{H} = \underbrace{\hbar \omega_0 \,\hat{a}^{\dagger} \hat{a}}_{\hat{H}_S}+\underbrace{\hbar g_1 (t) (\hat{a}\hat{b}_{1}^{\dagger}+ \hat{a}^{\dagger}\hat{b}_{1})+\hbar g_2 (t)(\hat{a}\hat{b}_{2}^{\dagger}+\hat{a}^{\dagger}\hat{b}_{2})}_{\hat{H}_{SB}}+\underbrace{\hat{H}_B}_{Reservoirs},
\end{equation}
where the decreasing time-dependent coupling functions are assumed as $g_1 (t)=\sqrt{\gamma_1}\,g(t)$, and $g_2 (t)=\sqrt{\gamma_2}\,g(t)$, and $\hat{H}_B$ is the Hamiltonian of the reservoirs to be determined later \cite{d22}.
The Hamiltonian Eq.~(\ref{1p}) can be rewritten as
\begin{equation}\label{3p}
      \hat{H} = \hbar \omega_0 \,\hat{a}^{\dagger} \hat{a}+\hbar\sqrt{\gamma_1+\gamma_2}\,g(t)\,[\hat{a}\,\hat{B}^\dag+\hat{a}^\dag\,\hat{B}]+\hbar\omega_0\hat{B}^\dag\hat{B},
\end{equation}
where the new reservoir operators are defined by
\begin{eqnarray}\label{newB}
     \hat{B} =& \frac{\sqrt{\gamma_1}\,\hat{b}_1+\sqrt{\gamma_2}\,\hat{b}_2}{\sqrt{\gamma_1+\gamma_2}},\nn\\
\hat{B}^\dag =& \frac{\sqrt{\gamma_1}\,\hat{b}^\dag_1+\sqrt{\gamma_2}\,\hat{b}^\dag_2}{\sqrt{\gamma_1+\gamma_2}},
\end{eqnarray}
and the Hamiltonian of the reservoirs is defined by
\begin{equation}\label{BathH}
\hat{H}_B=\hbar\omega_0\hat{B}^\dag\hat{B}.
\end{equation}
Now by using the Bogoliubov transformation
\begin{eqnarray}\label{BogoT}
\hat{C}=\frac{\hat{a}+\hat{B}}{\sqrt{2}},\nn\\
\hat{D}=\frac{\hat{B}-\hat{a}}{\sqrt{2}},
\end{eqnarray}
the Hamiltonian Eq.~(\ref{3p}) can be decoupled as
\begin{equation}\label{decoupled}
\hat{H}=\hbar(\omega_0+\sqrt{\gamma_1+\gamma_2}\,g(t))\,\hat{C}^\dag\hat{C}+\hbar(\omega_0-\sqrt{\gamma_1+\gamma_2}\,g(t))\,\hat{D}^\dag\hat{D}.
\end{equation}
From Heisenberg equations of motion we obtain
\begin{eqnarray}\label{4p}
&&  \hat{C}(t)=e^{-i[\omega t+\tilde{G}(t)]}\, \hat{C}(0),\nn\\
&&  \hat{D}(t)=e^{-i[\omega t-\tilde{G}(t)]}\, \hat{D}(0),
\end{eqnarray}
where for notational simplicity, we defined $\tilde{G}(t)=\sqrt{\gamma_1+\gamma_2}\,\int_0^t dt'\,g(t')$. Therefore, by making use of the inverse Bogoliuobov transformations, we finally obtain
\begin{equation}\label{5p}
  \begin{split}
      \hat{a} (t)= & e^{-i\omega_0 t}\Bigg\{\cos[\tilde{G}(t)]\hat{a} (0)-i \sin[\tilde{G}(t)]\Big(\frac{\sqrt{\gamma_1}}{\sqrt{\gamma_1+\gamma_2}}\,\hat{b}_1+\frac{\sqrt{\gamma_2}}{\sqrt{\gamma_1+\gamma_2}}\,\hat{b}_2\Big)\Bigg\},\\
      \hat{a}^\dag (t)= & e^{i\omega_0 t}\Bigg\{\cos[\tilde{G}(t)]\hat{a}^\dag (0)+i \sin[\tilde{G}(t)]\Big(\frac{\sqrt{\gamma_1}}{\sqrt{\gamma_1+\gamma_2}}\,\hat{b}^\dag_1+\frac{\sqrt{\gamma_2}}{\sqrt{\gamma_1+\gamma_2}}\,\hat{b}^\dag_2\Big)\Bigg\}.
  \end{split}
\end{equation}
The average number of bosons is
\begin{eqnarray}\label{6p}
 n(t) &=& \la\hat{a}^\dag (t)\hat{a} (t)\ra\nn\\
      &=& \cos^2[\tilde{G}(t)]\,\la\hat{a}^\dag (0)\hat{a} (0)\ra+\sin^2 [\tilde{G}(t)]\,\Big(\frac{\gamma_1}{\gamma_1+\gamma_2}\,\bar{n}_1+\frac{\gamma_2}{\gamma_1+\gamma_2}\,\bar{n}_2\Big),
\end{eqnarray}
where
\begin{eqnarray}\label{bathaverages}
&& \bar{n}_1=\frac{1}{e^{\frac{\hbar\omega_0}{k_{\mathrm{B}} T_1}}-1},\nn\\
&& \bar{n}_2=\frac{1}{e^{\frac{\hbar\omega_0}{k_{\mathrm{B}} T_2}}-1},
\end{eqnarray}
are the number densities of the left and right reservoirs, respectively. By choosing $\cos^2[\tilde{G}(t)]=e^{-\gamma t}$ \cite{d22}, and substituting it into Eq.~(\ref{6p}), we obtain
\begin{equation}\label{averagen}
 n(t)= \bar{n} +e^{-\gamma t}\,(n(0)-\bar{n}),
\end{equation}
where
\begin{equation}\label{averagen}
 \bar{n} =\frac{\gamma_1\,\bar{n}_1+\gamma_2\,\bar{n}_2}{\gamma_1+\gamma_2},
\end{equation}
is the steady state mean number. The energy of the oscillator at an arbitrary time is
\begin{eqnarray}\label{Energy}
  E(t) &=& \la \hbar\omega_0 \Big(\hat{a}^\dag(t)\hat{a}(t)+\frac{1}{2}\Big)\ra,\nn\\
       &=& \hbar\omega_0\,\Big(\bar{n} +e^{-\gamma t}\,(n(0)-\bar{n})+\frac{1}{2}\Big).
\end{eqnarray}
In the long-time regime, the energy tends to the stationary value $E(\infty)=E_s=\hbar\omega_0\,(\bar{n}+1/2)$, that is the oscillator has been thermalized to a bath with effective mean number $\bar{n}$.
\subsection{(Quasi)Distribution functions on the phase space of the main oscillator}
In quantum optics and quantum thermodynamics, quasi-probability distribution functions such as the Glauber-Sudarshan P-function, Wigner function, and Husimi Q-function are essential tools for representing quantum states in phase space. These functions provide insights into the quantum-classical boundary, coherence properties, and the non-classical nature of light.
\subsubsection{Husimi distribution function}\label{Usimi}
\begin{figure}[h]
\centering
\includegraphics[scale=0.8]{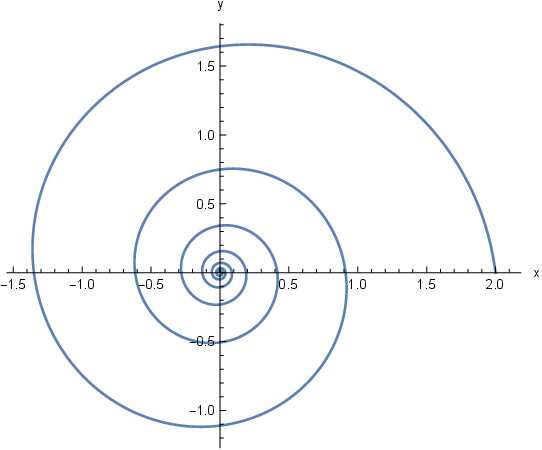}
\caption{(Color online) The locations of the maxima of the Husimi function (classical path) of the oscillator for the values $\bar{n}=0$, $\alpha_0=2$, and $\omega_0=4\gamma$ in terms of the dimensionless parameter $\tau=\gamma t$.}\label{fig3}
\end{figure}

The Husimi or $Q$-function corresponding to a density matrix $\hat{\rho}$ is a positive definite distribution on the phase space or complex plane defined by
\begin{equation}\label{Qfunction}
 Q(\alpha)=\frac{1}{\pi}\la \alpha|\hat{\rho}|\alpha\ra,
\end{equation}
where $|\alpha\ra$ is a coherent state. The Husimi distribution corresponding to the main oscillator with the reduced density matrix $\hat{\rho}_S (t)$ can be obtained as follows
\begin{equation}\label{Husimi}
\begin{split}
Q(\alpha,t)=& \frac{1}{\pi}\la\alpha|\hat{\rho}_S (t)|\alpha\ra,\\
           =& \frac{1}{\pi}\la 0|\hat{D}^\dag (\alpha)\hat{\rho}_S (t)\hat{D} (\alpha)|0\ra,\\
           =& \frac{1}{\pi} tr_{S}\,tr_{Baths}\,\big[|0\ra\la 0|\,\hat{D}^\dag (\alpha)\hat{\rho} (t)\hat{D} (\alpha)\big],\\
           =& \frac{1}{\pi} \sum_{s=0}^\infty \frac{(-1)^s}{s!}tr\,\big[\hat{U}^\dag (t)\hat{D} (\alpha)(\hat{a}^\dag)^s\hat{D}^\dag (\alpha)\hat{D} (\alpha)(\hat{a})^s\hat{D}^\dag (\alpha)\hat{U} (t)\hat{\rho} (0)\big],\\
\end{split}
\end{equation}
where $tr=tr_S\,tr_{Baths}$ is the total trace, $\hat{D} (\alpha)=e^{\alpha\hat{a}^\dag-\bar{\alpha}\hat{a}}$ is the displacement operator, and we made use of the identity \cite{Louisell}
\begin{equation}
|0\ra\la 0|=\sum_{s=0}^\infty \frac{(-1)^s}{s!}\,(\hat{a}^\dag)^s(\hat{a})^s,
\end{equation}
and the properties of the trace operator. Now using the identities
\begin{eqnarray}
\hat{D}(\alpha)\hat{a}^\dag \hat{D}(\alpha)=& \hat{a}^\dag-\bar{\alpha},\\
\hat{D}(\alpha)\hat{a}\hat{D}^\dag (\alpha)=& \hat{a}-\alpha,
\end{eqnarray}
and Heisenberg evolution of ladder operators
\begin{eqnarray}
\hat{U}^\dag (t)\hat{a} (0)\hat{U}(t)=& \hat{a} (t),\\
\hat{U}^\dag (t)\hat{a}^\dag (0)\hat{U}(t)=& \hat{a}^\dag (t),
\end{eqnarray}
Eq.~(\ref{Husimi}) can be rewritten as
\begin{equation}\label{simpleQ}
Q(\alpha,t)=\frac{1}{\pi}\sum_{s=0}^\infty \frac{(-1)^s}{s!}tr\,\big[(\hat{a}^\dag (t)-\bar{\alpha})^s\,(\hat{a} (t)-\alpha)^s\,\rho(0)\big].
\end{equation}
Let us assume that the main oscillator is initially prepared in a coherent state $\hat{\rho}_S (0)=|\alpha_0\ra\la \alpha_0|$, and the left and right bath oscillators are held in thermal states with inverse temperatures $\beta_{1(2)}$ and partition functions $z_{1(2)}$, respectively. Therefore,
\begin{equation}\label{initro}
\hat{\rho}(0)=|\alpha_0\ra\la \alpha_0|\otimes\underbrace{\frac{e^{-\beta_1\hbar\omega_0\,\hat{b}^\dag_1\hat{b}_1}}{z_1}\otimes\frac{e^{-\beta_2\hbar\omega_0\,\hat{b}^\dag_2\hat{b}_2}}{z_2}}_{\hat{\rho}_B (0)}.
\end{equation}
By inserting Eqs.~(\ref{5p}) and Eq.~(\ref{initro}) into Eq.~(\ref{simpleQ}), we find
\begin{equation}\label{Hubath}
\begin{split}
Q(\alpha,t)=& \frac{1}{\pi}\sum_{s=0}^\infty \frac{(-1)^s}{s!}\sum_{p,q=0}^s \binom{s}{p}\binom{s}{q}(\bar{\mu}\bar{\alpha}_0-\bar{\alpha})^{s-p}(\mu\alpha_0-\alpha)^{s-q}\\
           &\times tr_{b_1,b_2}\sum_{r=0}^p\sum_{u=0}^q (\bar{\nu}_1\,\hat{b}^\dag_1)^{p-r}(\bar{\nu}_2\,\hat{b}^\dag_2)^{r}(\nu_1\,\hat{b}_1)^{q-u}(\nu_2\,\hat{b}_2)^u\,\hat{\rho}_B (0),
\end{split}
\end{equation}
where for notational simplicity, we defined
\begin{equation}\label{defs}
\begin{split}
\mu(t)=& e^{-i\omega_0 t}\cos[\tilde{G}(t)],\\
\nu_1 =& -i \sqrt{\frac{\gamma_1}{\gamma_1+\gamma_2}}\,e^{-i\omega_0 t}\sin[\tilde{G}(t)],\\
\nu_2 =& -i \sqrt{\frac{\gamma_2}{\gamma_1+\gamma_2}}\,e^{-i\omega_0 t}\sin[\tilde{G}(t)].
\end{split}
\end{equation}
The bath terms in Eq.~(\ref{Hubath}) can be calculated using the identity
\begin{equation}
tr_b \Big((\hat{b}^\dag)^p(\hat{b})^q\,\frac{e^{-\beta\hbar\omega_0\,\hat{b}^\dag\hat{b}}}{z_b}\Big)=\delta_{pq}\,q!\,(\bar{n}_b)^q,
\end{equation}
where $\bar{n}_b=(e^{\beta\hbar\omega_0}-1)^{-1}$. Therefore, Eq.~(\ref{Hubath}) is more simplified as
\begin{equation}\label{HusimiL}
\begin{split}
Q(\alpha,t)=& \frac{1}{\pi}\sum_{s=0}^\infty \frac{(-1)^s}{s!}\sum_{p=0}^s \binom{s}{p}^2|\mu\alpha_0-\alpha|^{2(s-p)}\\
            & \times\sum_{u=0}^p \binom{p}{u}^2 (|\nu_1|^2\,\bar{n}_1)^p\Big(\frac{|\nu_2|^2\,\bar{n}_2}{|\nu_1|^2\,\bar{n}_1}\Big)^u (u!)(p-u)!\\
           =& \frac{1}{\pi}\sum_{s=0}^\infty \frac{(-1)^s}{s!}\sum_{p=0}^s \binom{s}{p}^2|\mu\alpha_0-\alpha|^{2(s-p)}\,p!\,\Big(|\nu_1|^2\,\bar{n}_1+|\nu_2|^2\,\bar{n}_2\Big)^p.
\end{split}
\end{equation}
Finally, using the following identities for Laguerre polynomials $L_s (x)$,
\begin{equation}\label{Laguerre}
\begin{split}
& L_s (x)=\sum_{u=0}^s \frac{s!}{(u!)^2 (s-u)!}(-x)^u,\\
& \sum_{s=0}^\infty z^s\,L_s (x)=\frac{e^{-\frac{zx}{1-z}}}{1-z},
\end{split}
\end{equation}
one obtains
\begin{eqnarray}\label{Hufinal}
Q(\alpha,t)=& \frac{1}{\pi}\,\frac{e^{-\frac{|\mu\alpha_0-\alpha|^2}{1+|\nu_1|^2\,\bar{n}_1+|\nu_2|^2\,\bar{n}_2}}}{1+\big[|\nu_1|^2\,\bar{n}_1+|\nu_2|^2\,\bar{n}_2\big]},\nn\\
           =& \frac{1}{\pi}\,\frac{e^{-\frac{|\mu\alpha_0-\alpha|^2}{1+\bar{n}\sin^2 [\tilde{G}(t)]}}}{1+\bar{n}\sin^2 [\tilde{G}(t)]},\nn\\
           =& \frac{1}{\pi}\,\dfrac{e^{-\dfrac{|\alpha-e^{-i\omega_0 t}e^{-\frac{\gamma t}{2}}\alpha_0|^2}{1+\bar{n}(1-e^{-\gamma t})}}}{1+\bar{n}(1-e^{-\gamma t})}.
\end{eqnarray}
Having the $Q$-function Eq.~(\ref{Hufinal}), we can obtain the expectations of anti-normal expressions, as an example, let us find $\la \hat{a}\hat{a}^\dag\ra$, we have
\begin{eqnarray}
  \la \hat{a}\hat{a}^\dag\ra &=& tr (\hat{\rho}(t)\hat{a}\hat{a}^\dag),\nn\\
                             &=& \int \frac{d^2\alpha}{\pi}\,\la \alpha|\hat{\rho}(t)\hat{a}\hat{a}^\dag|\alpha\ra,\nn \\
                             &=& \int \frac{d^2\alpha}{\pi}\,\la \alpha|\hat{a}^\dag\hat{\rho}(t)\hat{a}|\alpha\ra,\nn \\
                             &=& \int d^2\alpha\,|\alpha|^2\, Q(\alpha,t),\nn \\
                             &=& 1+|\alpha_0|^2 e^{-\gamma t}+\bar{n} (1-e^{-\gamma t}),
\end{eqnarray}
or equivalently $ \la \hat{a}^\dag\hat{a}\ra=|\alpha_0|^2 e^{-\gamma t}+\bar{n} (1-e^{-\gamma t})$. The locations of the maximum points of the Husimi distribution function in phase space lie on the curve $\alpha(t)=e^{-i\omega_0 t}e^{-\frac{\gamma t}{2}}\alpha_0$, which is a combination of rotation and contraction. In Fig.~(\ref{fig3}), the locations of the maxima of the Husimi function (classical path) has been depicted. The oscillator starts with the initial energy $E(0)=4.5\hbar\omega_0\,\,\,(\alpha_0=2)$ and tends to $\alpha=0$ with final energy $E(\infty)=0.5\hbar\omega_0$.
\subsubsection{Glauber-Sudarshan P-function}\label{pfunction}
The Glauber-Sudarshan or $P$-function expresses a quantum state's density matrix $\hat{\rho}$ as a weighted sum over coherent states $|\alpha\ra$
\begin{equation}\label{Pfunction}
  \hat{\rho}=\int d\alpha\,P(\alpha) |\alpha\ra\la\alpha|,
\end{equation}
This representation is particularly useful for distinguishing between classical and non-classical states. A well-behaved, positive $P$-function indicates a classical state, while a highly singular or negative $P$-function signifies non-classicality.

\begin{figure}[h]
\centering
\includegraphics[scale=0.8]{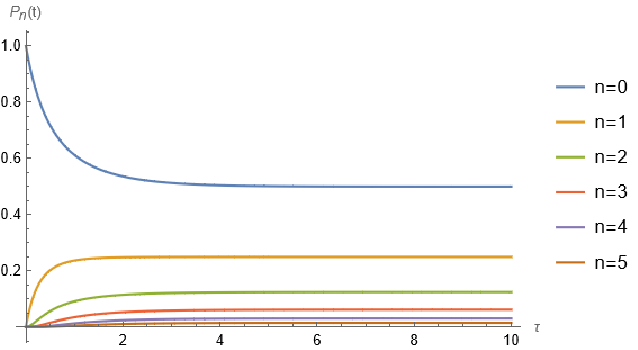}
\caption{(Color online) The probability function $P_n (t)$ in terms of the dimensionless parameter $\tau$ for $\bar{n}=1\,\,\, (\hbar\omega_0/k_B T\approx 0.693)$, and $n=0, 1, 2, 3, 4, 5$.}\label{fig4}
\end{figure}

The quantum optical characteristic functions corresponding to the reduced density matrix $\hat{\rho}_S (t)$ of the main oscillator are defined by \cite{GN}
\begin{eqnarray}\label{charac}
C_W (\lambda) &=& tr[\hat{\rho}_S (t)\,e^{\lambda\hat{a}^\dag-\bar{\lambda}\hat{a}}],\,\,\,\,\,\,\,(\text{Wigner})\\
C_N (\lambda) &=& tr[\hat{\rho}_S (t)\,e^{\lambda\hat{a}^\dag}e^{-\bar{\lambda}\hat{a}}],\,\,\,\,(\text{Normally ordered})\\
C_A (\lambda) &=& tr[\hat{\rho}_S (t)\,e^{-\bar{\lambda}\hat{a}}e^{\lambda\hat{a}^\dag}],\,\,\,\,(\text{Antinormally ordered}).
\end{eqnarray}
The characteristic functions are related as
\begin{equation}\label{relatedchara}
C_W (\lambda)=C_N (\lambda)e^{-\frac{1}{2}|\lambda|^2}=C_A (\lambda)e^{\frac{1}{2}|\lambda|^2}.
\end{equation}
The antinormally ordered characteristic function $C_A (\lambda)$ can be obtained from the Husimi distribution function as
\begin{equation}\label{CA}
  C_A (\lambda,t)=\int d^2 \alpha\,Q(\alpha,t)\,e^{\lambda\bar{\alpha}-\bar{\lambda}\alpha},
\end{equation}
therefore, by inserting Eq.~(\ref{Hufinal}) into Eq.~(\ref{CA}) we obtain
\begin{equation}\label{CA2}
 C_A (\lambda,t)=e^{-(1+\bar{n}\sin^2[\tilde{G}(t)])|\lambda|^2}\,e^{\lambda(\bar{\mu}\bar{\alpha}_0)-\bar{\lambda}(\mu\alpha_0)},
 \end{equation}
 therefore, using Eq.~(\ref{relatedchara}), we obtain
\begin{equation}\label{CA2}
 C_N (\lambda,t)=e^{-\bar{n}\sin^2[\tilde{G}(t)]|\lambda|^2}\,e^{\lambda(\bar{\mu}\bar{\alpha}_0)-\bar{\lambda}(\mu\alpha_0)}.
\end{equation}
Having the normally ordered characteristic function $C_N (\lambda)$, we can find the Glauber-Sudarshan $P$-function as
\begin{equation}\label{Pfunction}
\begin{split}
 P(\alpha,t)=& \frac{1}{\pi^2}\int d^2\lambda\,C_N (\lambda,t)\,e^{\bar{\lambda}\alpha-\lambda\bar{\alpha}},\\
            =& \frac{1}{\pi^2}\int d^2\lambda\, e^{-\bar{\lambda}(\bar{n}\sin^2[\tilde{G}(t)])\lambda+\lambda(\bar{\mu}\bar{\alpha}_0-\bar{\alpha})+\bar{\lambda}(\alpha-\mu\alpha_0)},\\
            =& \frac{1}{\pi\bar{n}\sin^2[\tilde{G}(t)]}\,e^{-\frac{|\alpha-\mu\alpha_0|^2}{\bar{n}\sin^2[\tilde{G}(t)]}},\nn\\
            =& \frac{1}{\pi\bar{n}(1-e^{-\gamma t})}\,e^{-\dfrac{|\alpha-e^{-i\omega_0 t}e^{-\frac{\gamma t}{2}}\alpha_0|^2}{\bar{n}(1-e^{-\gamma t})}}.
\end{split}
\end{equation}
The probability of finding the oscillator in the number state $|n\ra$ at an arbitrary time when it is initially prepared in a coherent state $|\alpha_0\ra\la \alpha_0|$, can be calculated as follows
\begin{eqnarray}\label{Pnt}
  P_n (t) &=& \int d\alpha\,P(\alpha,t)\la n|\alpha\ra\la n|\alpha\ra,\nn \\
          &=& \frac{1}{\pi\sigma n!} \int d\alpha\,e^{-\frac{1}{\sigma}(\bar{\alpha}-\bar{\delta})(\alpha-\delta)} e^{-|\alpha|^2} \alpha^n \bar{\alpha}^n,\nn\\
          &=& \frac{e^{-\frac{|\delta|^2}{\sigma}}}{\sigma n!}\frac{\partial^n}{\partial\bar{\lambda}^n}\frac{\partial^n}{\partial\lambda^n}\int d\alpha\,e^{-\bar{\alpha}(\frac{1+\sigma}{\sigma})\alpha+(\frac{\delta}{\sigma}+\lambda)\bar{\alpha}+
       (\frac{\bar{\delta}}{\sigma}+\bar{\lambda})\alpha}\Bigg|_{\lambda=\bar{\lambda}=0},\nn\\
          &=& \frac{\sigma^n}{(1+\sigma)^{n+1}}\,e^{-\frac{|\delta|^2}{1+\sigma}}\,L_n \Big(-\frac{|\delta|^2}{\sigma(1+\sigma)}\Big),
\end{eqnarray}
where for notational simplicity we defined
\begin{eqnarray}
  \sigma &=& \bar{n} (1-e^{-\gamma t}),\nn \\
  \delta &=& e^{-i\omega_0 t}e^{-\frac{\gamma t}{2}}\alpha_0,
\end{eqnarray}
and made use of the definition of Laguerre polynomials
\begin{equation}\label{Laguerre}
  L_n (x)=\sum_{s=0}^n {{n}\choose{s}} \frac{(-1)^s}{s!} x^s.
\end{equation}
If the baths are held in zero temperature, $(\bar{n}=0)$, Eq.~(\ref{Pnt}) becomes
\begin{equation}\label{zerotemp}
  P^{\tiny T=0}_n (t)=\frac{e^{-|\delta|^2}\,|\delta|^{2n}}{n!},
\end{equation}
which is a Poisson distribution with mean number $\la n\ra=|\delta|^2=e^{-\gamma t} |\alpha_0|^2$. In the long-time regime, $P_n (t)$ tends to the stationary distribution
\begin{equation}\label{stationaryTh}
  P_n (\infty)=\frac{\bar{n}^n}{(1+\bar{n})^{n+1}},
\end{equation}
which is a thermal distribution as expected from thermalization condition. In Fig.~(\ref{fig4}), the probability function $P_n (t)$ has been depicted for various $n$ in terms of the dimensionless parameter $\tau$.

\begin{figure}[h]
\centering
\includegraphics[scale=0.8]{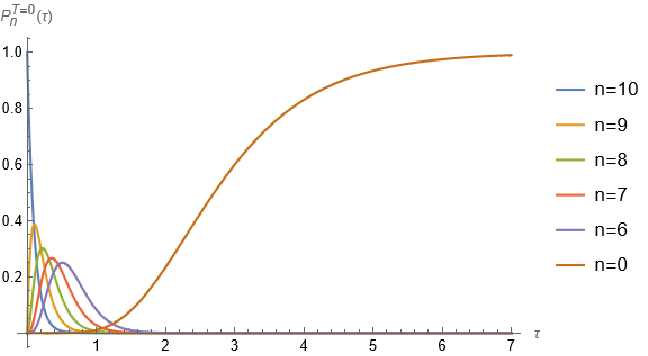}
\caption{(Color online) The probability function $P^{T=0}_n (t)$ in terms of the dimensionless parameter $\tau$ for $\bar{n}=0$, and $n=0,6,7,8,9,10$.}\label{fig5}
\end{figure}

Now let the initial state of the oscillator be a number state $|N\ra\la N|$, then
\begin{equation}\label{Initnum}
  \hat{\rho}(0)=|N\ra\la N|\otimes\underbrace{\frac{e^{-\beta_1\hbar\omega_0\,\hat{b}^\dag_1\hat{b}_1}}{z_1}\otimes\frac{e^{-\beta_2\hbar\omega_0\,\hat{b}^\dag_2\hat{b}_2}}{z_2}}_{\hat{\rho}_B (0)},
\end{equation}
and by making use of the formula \cite{d22}
\begin{equation}\label{elegantformula}
  \la n|\hat{\rho}^S (t)|m\ra=\frac{1}{\sqrt{n! m!}}\sum_{s=0}^\infty \frac{(-1)^s}{s!}\,tr\Big[\big(\hat{a}^\dag (t)\big)^{s+m}\big(\hat{a}(t)\big)^{s+n}\,\hat{\rho}(0)\Big],
\end{equation}
one deduces that the reduced density matrix $\hat{\rho}^S (t)$ is diagonal in number state basis and therefore, there will be no decoherency. When the bath oscillator is held in zero temperature, one easily finds
\begin{equation}\label{Numberstateprob}
  P_n^{T=0}(t)=\left\{
              \begin{array}{ll}
                {{n}\choose{N}}(\cos^2[\tilde{G}(t)])^n\,(\sin^2[\tilde{G}(t)])^{N-n}, & n\leq N, \\
                0, & n>N,
              \end{array}
            \right.
\end{equation}
which is a binomial distribution or a random walk with right and left probabilities given by $p=\cos^2[\tilde{G}(t)]$, and $q=\sin^2[\tilde{G}(t)]$, respectively. In Fig.~(\ref{fig5}) the probability function $ P_n^{T=0}(t)$ has been depicted in terms of the dimensionless parameter $\tau$ for different values of $n$ and $\cos^2[\tilde{G}(t)]=e^{-\gamma t}$.
\subsubsection{Wigner function}\label{Wf}
\subsubsection{The initial state is a coherent state}
For a coherent state $\hat{\rho}^S (0)=|\alpha_0\ra\la \alpha_0|$, the normally ordered characteristic function $C_N(\lambda,t)$ is given in Eq.~(\ref{CA2}), and the corresponding Wigner function can be obtained as
\begin{eqnarray}\label{Wigner}
  W_{coh}(\alpha,t) &=& \frac{1}{\pi^2}\,\int d\lambda\,C_N(\lambda,t)\,e^{-\frac{|\lambda|^2}{2}}\,e^{\bar{\lambda}\alpha-\lambda\bar{\alpha}},\nn\\
             &=& \frac{1}{\pi^2}\,\int d\lambda\,e^{-\lambda\big(\frac{1}{2}+\bar{n}\sin^2[\tilde{G}(t)]\big)\lambda+\lambda(\bar{\mu}\bar{\alpha}_0-\bar{\alpha})+\bar{\lambda}(\alpha-\mu\alpha_0)},\nn\\
             &=& \frac{1}{\pi}\,\frac{1}{\frac{1}{2}+\bar{n}(1-e^{-\gamma t})}\,e^{-\dfrac{|\alpha-e^{-i\omega_0 t}e^{-\frac{\gamma t}{2}}\alpha_0|^2}{\frac{1}{2}+\bar{n}(1-e^{-\gamma t})}},
\end{eqnarray}
which is a positive function, as expected from a coherent state.
\subsubsection{The initial state is a number state}
\begin{figure}[H]
    \centering
    \subfigure[]{\includegraphics[width=0.3\textwidth]{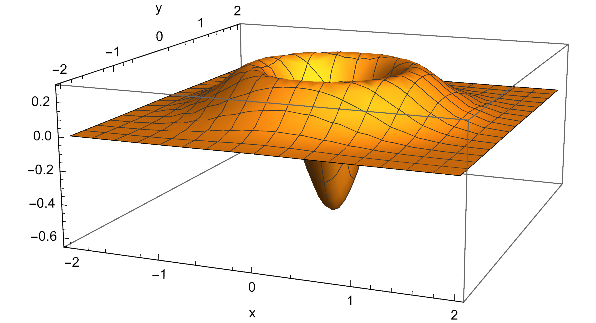}}
    \subfigure[]{\includegraphics[width=0.3\textwidth]{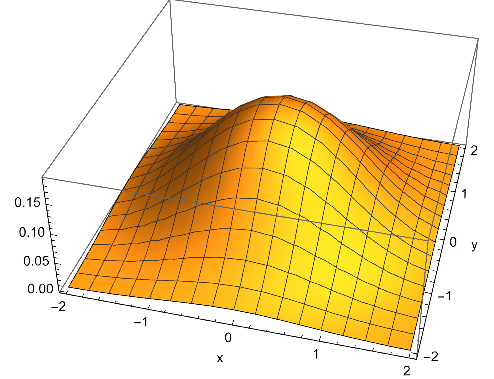}}
       \caption{(a) $N=1,\,\,\, \tau=0,\,\bar{n}=1$ (b) $N=1,\,\,\, \tau=1,\,\bar{n}=1$.}
    \label{fig6}
\end{figure}

For a number state $\hat{\rho}^S (0)=|N\ra\la N|$, The normally ordered characteristic function $C^{num}_N(\lambda,t)$ is calculated as follows
\begin{eqnarray}\label{CNnumber}
  C^{num}_N(\lambda,t) &=& tr\Big[\hat{\rho}^S (t)\,e^{\lambda\hat{a}^\dag} e^{-\bar{\lambda}\hat{a}}\Big],\nn \\
                       &=& tr\Big[\hat{\rho}^S (0)\,e^{\lambda\hat{a}^\dag}(t) e^{-\bar{\lambda}\hat{a}(t)}\Big],\nn \\
                       &=& \la N|e^{\lambda\bar{\mu}\hat{a}^\dag}e^{-\bar{\lambda}\mu\hat{a}}|N\ra tr\Big[\hat{\rho}_B (0)\,e^{\lambda\bar{\nu}\hat{B}^\dag}e^{\bar{-\lambda}\nu\hat{B}}\Big],\nn\\
                       &=& L_{N}(|\lambda|^2\cos^2[\tilde{G}])\,e^{-\bar{n}\sin^2[\tilde{G}] |\lambda|^2},
\end{eqnarray}
where we made use of Eqs.~(\ref{Laguerre},\ref{initro}). Therefore, for the corresponding Wigner function we have
\begin{eqnarray}\label{WignerNum}
 W_{N}(\alpha,t) &=& \frac{1}{\pi^2}\int d^2\lambda\, C^{num}_N(\lambda,t)\,e^{-\frac{1}{2}|\lambda|^2}\,e^{\bar{\lambda}\alpha-\lambda\bar{\alpha}},\nn \\
                         &=& \frac{1}{\pi^2}\int d^2\lambda\,L_{N}(|\lambda|^2\cos^2[\tilde{G}])\,e^{-\bar{n}\sin^2[\tilde{G}] |\lambda|^2}\,e^{-\frac{1}{2}|\lambda|^2}\,e^{\bar{\lambda}\alpha-\lambda\bar{\alpha}},
\end{eqnarray}
where we made use of the identity $\sum\limits_{s=0}^\infty y^s\,L_s (x)=\frac{e^{-\frac{xy}{1-y}}}{1-y}$. Now, by multiplying both sides of Eq.~(\ref{WignerNum}) by $y^N$ and adding over $N$, we have for the generating function
\begin{eqnarray}\label{GenW}
 \sum_{N=0}^\infty y^N\,W_{N}(\alpha,t) &=& \frac{1}{\pi^2}\int d^2\lambda\,\frac{1}{1-y} e^{-\bar{\lambda}\big[\frac{y\,\cos^2[\tilde{G}]}{1-y}+\bar{n}\sin^2[\tilde{G}]\big]\lambda+\frac{1}{2}\lambda+\bar{\lambda}\alpha-\lambda\bar{\alpha}},\nn\\
   &=& \frac{1}{\pi^2} \frac{1}{[y\cos^2[\tilde{G}]+(\bar{n}\sin^2[\tilde{G}]+\frac{1}{2})(1-y)]}\,e^{-\frac{(1-y)|\alpha|^2}{[y\cos^2[\tilde{G}]+(\bar{n}\sin^2[G]+\frac{1}{2})(1-y)]}}.
\end{eqnarray}
From the generating function Eq.~(\ref{GenW}), we obtain the Wigner function as
\begin{equation}\label{WignerNumber}
  W_{N}(\alpha,t)=\frac{(-1)^N}{\pi}\frac{(\psi(t))^N}{(\phi(t))^{N+1}}\, e^{-\frac{|\alpha|^2}{\phi(t)}}\,L_N \Big(\frac{\phi(t)+\psi(t)}{\phi(t)\psi(t)}\,|\alpha|^2\Big),
\end{equation}
where for notational simplicity we defined
\begin{eqnarray}\label{phipsi}
  \phi(t) &=& \bar{n}\sin^2 [\tilde{G}]+\frac{1}{2},\nn \\
  \psi(t) &=& \cos^2[\tilde{G}]-\bar{n}\sin^2 [\tilde{G}]-\frac{1}{2}.
\end{eqnarray}
In Fig.~(\ref{fig6}), the Wigner function Eq.~(\ref{WignerNumber}) has been depicted for $\tau=0, 1,\,\,\,\bar{n}=1$, and $\cos^2[\tilde{G}(t)]=e^{-\gamma t}$ in terms of the dimensionless parameter $\tau=\gamma t$. The Wigner function is negative at $\tau=0$ corresponding to the nonclassical number state $|1\ra$, and it becomes positive at $\tau=1$ where the system state tends to the coherent state $|0\ra$.
In the long-time regime ($t\gg \gamma^{-1}$), the Wigner function Eq.~(\ref{WignerNumber}) tends to the thermalized Wigner state
\begin{equation}\label{WigNumTher}
   W_{N}(\alpha,\infty)=\frac{1}{\pi}\frac{1}{\bar{n}+\frac{1}{2}}\,e^{-\frac{|\alpha|^2}{\bar{n}+\frac{1}{2}}},
\end{equation}
corresponding to the ground state $|0\ra\la 0|$, as expected.
\section{Generalisation: An oscillator interacting with an arbitrary number of thermal reservoirs with different temperatures}\label{Generalization}
\begin{figure}[h]
\centering
\includegraphics[scale=0.15]{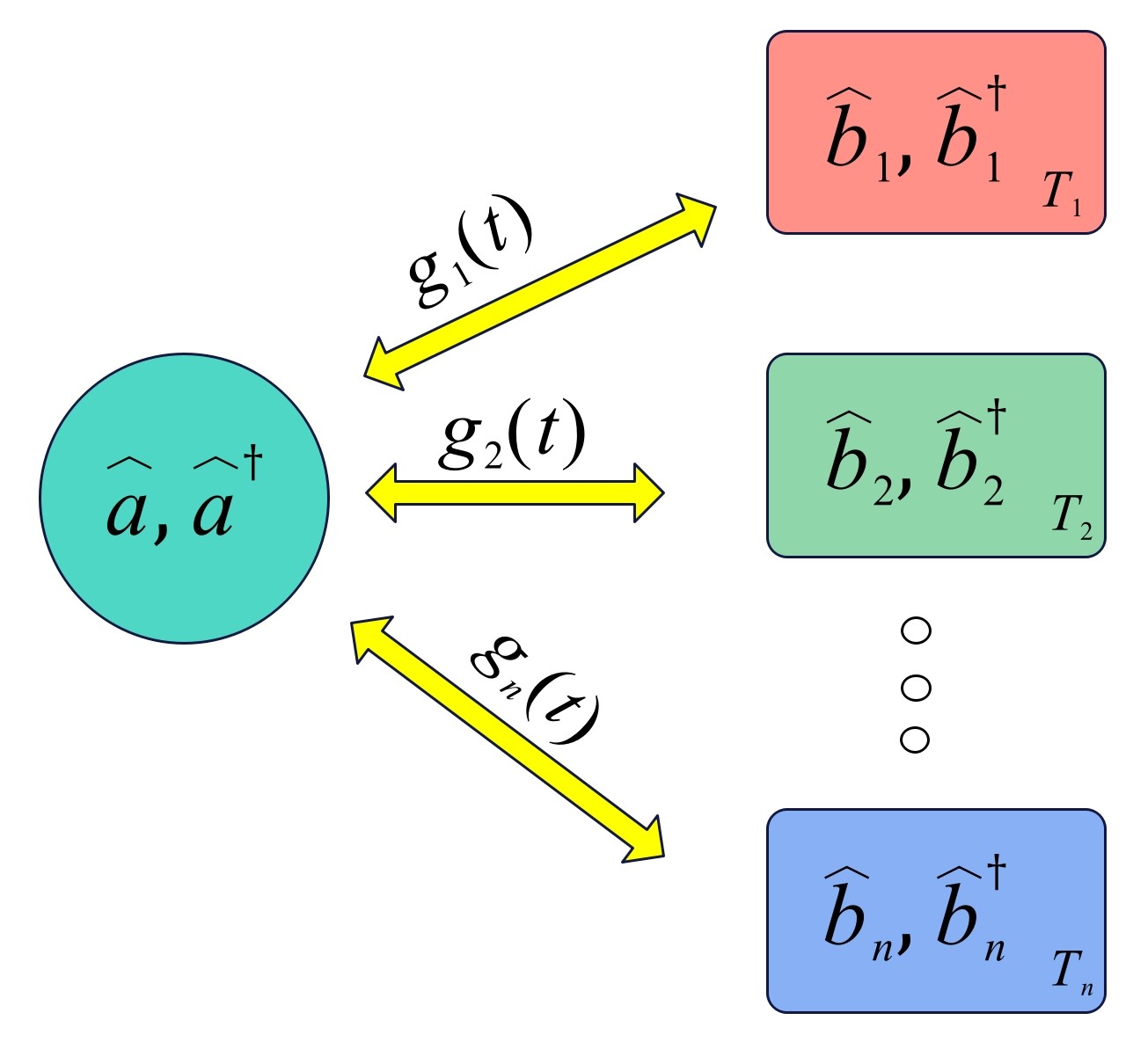}
\caption{(Color online) An oscillator interacting with n thermal reservoirs with temperatures $T_1$, $T_2$,..., and $T_n$.}\label{fig7}
\end{figure}

Let the main oscillator be interacting with $n$ reservoirs with temperatures $T_1$, $T_2$,...,and $T_n$ trough the time-dependent coupling functions $g_k (t)=\sqrt{\gamma_k} g(t)$, Fig.~(\ref{fig7}). The total Hamiltonian can be written as
\begin{eqnarray}\label{M1}
\hat{H} &=& \underbrace{\hbar \omega_0 \,\hat{a}^{\dagger} \hat{a}}_{\hat{H}_S}+\underbrace{\sum_{k=1}^n\hbar g_k (t) (\hat{a}\hat{b}_{k}^{\dagger}+ \hat{a}^{\dagger}\hat{b}_{k})}+\underbrace{\hat{H}_B}_{Reservoirs},\nn\\
        &=& \hbar \omega_0 \,\hat{a}^{\dagger} \hat{a}+\hbar\sqrt{\gamma}\,g(t)[\hat{a}\hat{B}^\dag+\hat{a}^\dag\hat{B}]+\hbar\omega_0\hat{B}^\dag\hat{B},
\end{eqnarray}
where $\gamma=\sum\limits_{k=1}^n \gamma_k$, and the Bath operators are defined by
\begin{eqnarray}\label{M2}
     \hat{B} &=& \frac{\sum\limits_{k=1}^n\sqrt{\gamma_k}\,\hat{b}_k}{\sqrt{\gamma}},\nn\\
     \hat{B}^\dag &=& \frac{\sum\limits_{k=1}^n\sqrt{\gamma_k}\,\hat{b}^\dag_k}{\sqrt{\gamma}}.
\end{eqnarray}
The Hamiltonian Eq.~(\ref{M1}) is like the Hamiltonian Eq.~(\ref{1p}), therefore, following the same steps, one finds
\begin{equation}\label{M3}
  \begin{split}
      \hat{a} (t)= & e^{-i\omega_0 t}\Bigg\{\cos[\tilde{G} (t)]\hat{a} (0)-i \sin[\tilde{G} (t)]\Bigg(\frac{\sum\limits_{k=1}^n\sqrt{\gamma_k}\,\hat{b}_k}{\sqrt{\gamma}}\Bigg)\Bigg\},\\
      \hat{a}^\dag (t)= & e^{i\omega_0 t}\Bigg\{\cos[\tilde{G} (t)]\hat{a}^\dag (0)+i \sin[\tilde{G} (t)]\Bigg(\frac{\sum\limits_{k=1}^n\sqrt{\gamma_k}\,\hat{b}^\dag_k}{\sqrt{\gamma}}\Bigg)\Bigg\},
  \end{split}
\end{equation}
where we defined $\tilde{G} (t)=\sqrt{\gamma}\,\int_0^t dt'\,g(t')$. For the initial state
\begin{equation}\label{M4}
  \hat{\rho} (0)=\hat{\rho}_S\otimes\underbrace{\frac{e^{-i\hbar\omega_0\hat{b}^\dag_1\hat{b}_1}}{z_{b_1}}}_{\hat{\rho}_1 (0)}\otimes\cdots\otimes\underbrace{\frac{e^{-i\hbar\omega_0\hat{b}^\dag_n\hat{b}_n}}{z_{b_n}}}_{\hat{\rho}_n (0)},
\end{equation}
the mean number of excitations is
\begin{eqnarray}\label{M5}
n(t) &=& \cos^2 [\tilde{G} (t)]\,n(0)+\sin^2 [\tilde{G} (t)]\underbrace{\sum_{k=1}^n\frac{\gamma_k \bar{n}_k}{\gamma}}_{\bar{n}},\nn\\
     &=& e^{-\gamma t}\,n(0)+(1-e^{-\gamma t})\,\bar{n},
\end{eqnarray}
where we used  $\cos^2 [\tilde{G} (t)]=e^{-\gamma t}$, and $\bar{n}_k=tr_k [\hat{\rho}_k (0)\hat{b}^\dag_k\hat{b}_k]$. Note that from long-time behavior and thermalization conditions, one easily finds $\gamma=\sum\limits_{k=1}^n \gamma_k$. Therefore, an oscillator interacting with $n$ oscillators with different temperatures and coupling strengths $\sqrt{\gamma_k}$ is equivalent to an oscillator interacting with a thermal bath with effective mean number
$\bar{n}=\frac{1}{\gamma}\sum\limits_{k=1}^n \gamma_k \bar{n}_k$ and the effective strength $\sqrt{\gamma}$.
\subsection{Quantum current}\label{current}
From the definition of the effective mean number
\begin{equation}\label{nbar}
  \bar{n}=\frac{1}{\gamma}\sum\limits_{k=1}^n \gamma_k \bar{n}_k,\,\,\,(\gamma=\sum_{k=1}^n \gamma_k),
\end{equation}
we find
\begin{equation}\label{I1}
  \sum_{k=1}^n \gamma_k(\bar{n}_k-\bar{n})=0,
\end{equation}
indicating that in the stationary state (long-time regime), the algebraic sum of currents flowing in and out of the oscillator is zero. Therefore, the sum of currents flowing into the oscillator from reservoirs ($I^{in}_s$) is equal to the sum of currents flowing from the oscillator into the reservoirs ($I^{out}_s$). In the finite time, the quantum current can be defined as the average of the input-output currents by replacing $\bar{n}$ with $n(t)$ \cite{c1}, so we define
\begin{equation}\label{It}
  I(t)=\frac{1}{2}\sum_{k=1}^n \gamma_k|\bar{n}_k-n(t)|.
\end{equation}
The stationary current is
\begin{equation}\label{StationaryI}
I_s=I(\infty)=\frac{1}{2}\sum_{k=1}^n \gamma_k|\bar{n}_k-\bar{n}|.
\end{equation}
\subsubsection{Example}
For three baths ($n=3$), with equal couplings $\gamma_1=\gamma_2=\gamma_3=1$, and corresponding thermal densities $\bar{n}_1=5,\,\bar{n}_2=2,\,\bar{n}_3=5$, we have $\bar{n}=2$. For the main oscillator let us assume $n(0)=5$, then $n(t)=4+e^{-3t}$, and the current is
\begin{eqnarray}\label{current3}
  I(t) &=& \frac{1}{2}\Big(|5-4-e^{-3t}|+|2-4-e^{-3t}|+|5-4-e^{-3t}|\Big),\nn\\
       &=& 2-\frac{1}{2}\,e^{-3t}.
\end{eqnarray}
\section{A Two-level system interacting with two thermal reservoirs}\label{Sec2}
\begin{figure}[h]
\centering
\includegraphics[scale=0.15]{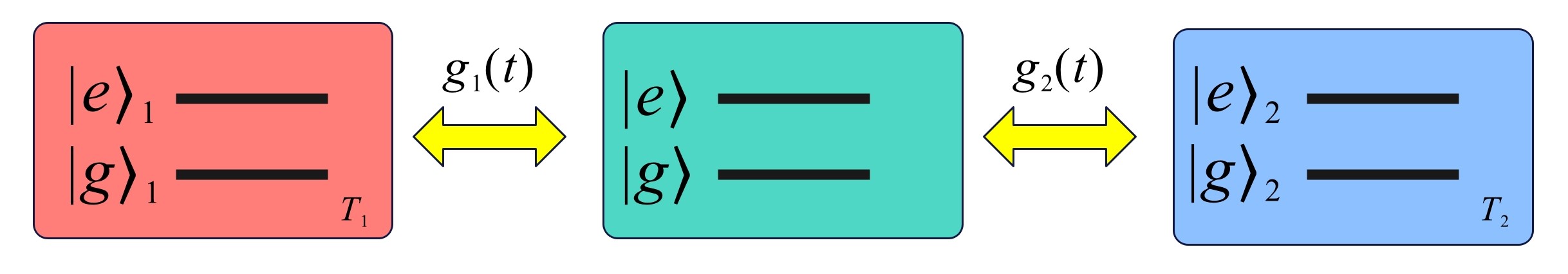}
\caption{ (Color online) A two-level system interacting with two thermal reservoirs with temperatures $T_1$ and $T_2$, and time-dependent coupling functions $g_1 (t)=\sqrt{\gamma_1}\,g(t)$ and $g_2 (t)=\sqrt{\gamma_2}\,g(t)$.}\label{fig8}
\label{fig:1}
\end{figure}

Examining two-level systems interacting with reservoirs at different temperatures enhances our understanding of quantum energy transfer mechanisms, which is essential for designing efficient energy-harvesting technologies. Additionally, the coupling of two-level systems with thermal reservoirs at varying temperatures offers valuable insights into decoherence processes, playing a key role in preserving quantum coherence in quantum computing \cite{ban2019decoherence}. In this section, we investigate the dynamics of a two-level system interacting with two reservoirs with temperatures $T_1$ and $T_2$ trough time-dependent coupling functions $g_1 (t)=\sqrt{\gamma_1}\,g(t)$ and $g_2 (t)=\sqrt{\gamma_2}\,g(t)$, respectively, Fig.~(\ref{fig8}).
The Hamiltonian of the system is
\begin{equation}\label{b1}
  \begin{split}
     H(t)= & \frac{\hbar \omega_0}{2}(\sigma^{1}_{z}+\sigma^{2}_{z}+\sigma_{z} )+\hbar \sqrt{\gamma_{1}}g(t)(\sigma^{1}_{-}\sigma_{+}+\sigma^{1}_{+}\sigma_{-})\\
       & +\hbar \sqrt{\gamma_{2}}g(t)(\sigma^{2}_{-}\sigma_{+}+\sigma^{2}_{+}\sigma_{-}).
  \end{split}
\end{equation}
In the standard basis the state of the total system can be expanded as
\begin{equation}\label{b2}
  |\psi(t)\ra=\sum_{i_1,i,i_2\in\{+,-\}}\,C_{i_1,i,i_2} (t)|i_1\ra\otimes|i\ra\otimes|i_2\ra.
\end{equation}
By inserting Eq.~(\ref{b2}) into the Schr\"{o}dinger equation
\begin{equation}\label{SchroEq}
 i\hbar\frac{d}{dt} |\psi(t)\ra=\hat{H}(t) |\psi(t)\ra,
\end{equation}
we find the following set of equations
\begin{eqnarray}
 i\hbar\frac{d}{dt}C_{+++}(t)&=&\frac{3}{2}\hbar\omega_0\,C_{+++} (t),\\
 i\hbar\frac{d}{dt}C_{---}(t)&=&-\frac{3}{2}\hbar\omega_0\,C_{---} (t),
\end{eqnarray}
with solutions
\begin{eqnarray}
  C_{+++}(t) &=& e^{-\frac{3}{2}i\omega_0 t}\,C_{+++}(0), \\
  C_{---}(t) &=& e^{\frac{3}{2}i\omega_0 t}\,C_{---}(0),
\end{eqnarray}
and two independent sets of matrix equations
\begin{equation}\label{b12}
 i \hbar \begin{pmatrix}
    \dot{C}_{++-}(t) \\
    \dot{C}_{+-+}(t) \\
    \dot{C}_{-++}(t)
  \end{pmatrix}=
 \underbrace{ \begin{pmatrix}
    \frac{\hbar \omega_0 }{2} & \hbar \sqrt{\gamma_2}g(t) & 0 \\
    \hbar \sqrt{\gamma_2}g(t)&\frac{\hbar \omega_0 }{2} & \hbar \sqrt{\gamma_1}g(t) \\
    0 & \hbar \sqrt{\gamma_1}g(t) & \frac{\hbar \omega_0 }{2}
  \end{pmatrix}}_{\hbar \Lambda (t)}
  \begin{pmatrix}
     {C}_{++-}(t)  \\
     {C}_{+-+}(t)  \\
     {C}_{-++}(t)
   \end{pmatrix},
\end{equation}
\begin{equation}\label{b12}
 i \hbar \begin{pmatrix}
    \dot{C}_{+--}(t) \\
    \dot{C}_{-+-}(t) \\
    \dot{C}_{--+}(t)
  \end{pmatrix}=
 \underbrace{ \begin{pmatrix}
   - \frac{\hbar \omega_0 }{2} & \hbar \sqrt{\gamma_1}g(t) & 0 \\
    \hbar \sqrt{\gamma_1}g(t)&-\frac{\hbar \omega_0 }{2} & \hbar \sqrt{\gamma_2}g(t) \\
    0 & \hbar \sqrt{\gamma_2}g(t) & -\frac{\hbar \omega_0 }{2}
  \end{pmatrix}}_{\hbar \Gamma(t)}
  \begin{pmatrix}
     {C}_{+--}(t)  \\
     {C}_{-+-}(t)  \\
     {C}_{--+}(t)
   \end{pmatrix},
\end{equation}
The matrices $\Lambda(t)$ and $\Gamma(t)$ commute at different times $[\Lambda(t),\Lambda(t')]=[\Gamma(t),\Gamma(t')]=0$, therefore, the matrix equations can be solved by calculating the evolution matrices $M=e^{-i\int^{t}_{0}\Lambda(t')dt'}$, and $N=e^{-i\int^{t}_{0}\Gamma(t')dt'}$, as
\begin{equation}\label{b13}
  \begin{pmatrix}
     {C}_{+--}(t)  \\
     {C}_{-+-}(t)  \\
     {C}_{--+}(t)
   \end{pmatrix}=e^{-i\int^{t}_{0}\Gamma(t')dt'}\begin{pmatrix}
     {C}_{+--}(0)  \\
     {C}_{-+-}(0)  \\
     {C}_{--+}(0)
   \end{pmatrix},
\end{equation}
\begin{equation}\label{b13}
  \begin{pmatrix}
     {C}_{++-}(t)  \\
     {C}_{+-+}(t)  \\
     {C}_{-++}(t)
   \end{pmatrix}=e^{-i\int^{t}_{0}\Lambda(t')dt'}\begin{pmatrix}
     {C}_{++-}(0)  \\
     {C}_{+-+}(0)  \\
     {C}_{-++}(0)
   \end{pmatrix}.
\end{equation}
The matrix exponentials can be found using the Wolfram Mathematica \cite{ram2022}, therefore,
\begin{equation}\label{b14}
 \begin{pmatrix}
     {C}_{++-}(t)  \\
     {C}_{+-+}(t)  \\
     {C}_{-++}(t)
   \end{pmatrix}=\underbrace{\begin{pmatrix}
                   \frac{e^{-i\omega_0 t/2}[\gamma_1+\gamma_2 \cos \tilde{G}]}{\gamma_1+\gamma_2} & \frac{-ie^{-i\omega_0 t/2}\sqrt{\gamma_2}\sin\tilde{G}}{\sqrt{\gamma_1+\gamma_2}} & \frac{e^{-i\omega_0 t/2}\sqrt{\gamma_1 \gamma_2}[\cos \tilde{G}-1]}{\gamma_1+\gamma_2} \\
                   \frac{-ie^{-i\omega_0 t/2}\sqrt{\gamma_2}\sin\tilde{G}}{\sqrt{\gamma_1+\gamma_2}} & e^{-i\omega_0 t/2} \cos \tilde{G}  & \frac{-ie^{-i\omega_0 t/2}\sqrt{\gamma_1}\sin\tilde{G}}{\sqrt{\gamma_1+\gamma_2}} \\
                    \frac{e^{-i\omega_0 t/2}\sqrt{\gamma_1 \gamma_2}[\cos \tilde{G}-1]}{\gamma_1+\gamma_2} & \frac{-ie^{-i\omega_0 t/2}\sqrt{\gamma_1}\sin\tilde{G}}{\sqrt{\gamma_1+\gamma_2}} & \frac{e^{-i\omega_0 t/2}[\gamma_2+\gamma_1 \cos \tilde{G}]}{\gamma_1+\gamma_2}
                 \end{pmatrix}}_{N} \begin{pmatrix}
     {C}_{++-}(0)  \\
    {C}_{+-+}(0)  \\
     {C}_{-++}(0)
   \end{pmatrix},
\end{equation}
\begin{equation}\label{b15}
 \begin{pmatrix}
     {C}_{+--}(t)  \\
    {C}_{-+-}(t)  \\
     {C}_{--+}(t)
   \end{pmatrix}=\underbrace{\begin{pmatrix}
                   \frac{e^{i\omega_0 t/2}[\gamma_2+\gamma_1 \cos \tilde{G}]}{\gamma_1+\gamma_2} & \frac{-ie^{i\omega_0 t/2}\sqrt{\gamma_1}\sin\tilde{G}}{\sqrt{\gamma_1+\gamma_2}} & \frac{e^{i\omega_0 t/2}\sqrt{\gamma_1 \gamma_2}[\cos \tilde{G}-1]}{\gamma_1+\gamma_2} \\
                   \frac{-ie^{i\omega_0 t/2}\sqrt{\gamma_1}\sin\tilde{G}}{\sqrt{\gamma_1+\gamma_2}} & e^{i\omega_0 t/2} \cos \tilde{G}  & \frac{-ie^{i\omega_0 t/2}\sqrt{\gamma_2}\sin\tilde{G}}{\sqrt{\gamma_1+\gamma_2}} \\
                    \frac{e^{i\omega_0 t/2}\sqrt{\gamma_1 \gamma_2}[\cos \tilde{G}-1]}{\gamma_1+\gamma_2} & \frac{-ie^{i\omega_0 t/2}\sqrt{\gamma_2}\sin\tilde{G}}{\sqrt{\gamma_1+\gamma_2}} & \frac{e^{i\omega_0 t/2}[\gamma_1+\gamma_2 \cos \tilde{G}]}{\gamma_1+\gamma_2}
                 \end{pmatrix} }_{M}\begin{pmatrix}
     {C}_{+--}(0)  \\
    {C}_{-+-}(0)  \\
     {C}_{--+}(0)
   \end{pmatrix},
\end{equation}
where $\tilde{G}(t)=\sqrt{\gamma_1+\gamma_2}\int^{t}_{0}g(t')dt'$, and matrices $M$ and $N$ are defined for notational simplicity. In the standard basis, the evolution equation can be written as
\begin{equation}\label{evolution}
\left(
  \begin{array}{c}
    C_{+++}(t) \\
    C_{++-}(t) \\
    C_{+-+}(t) \\
    C_{+--}(t) \\
    C_{-++}(t) \\
    C_{-+-}(t) \\
    C_{--+}(t) \\
    C_{---}(t) \\
  \end{array}
\right)=\underbrace{\left(
          \begin{array}{cccccccc}
            e^{-\frac{3}{2}i\omega_0 t} & 0 & 0 & 0 & 0 & 0 & 0 & 0 \\
            0 & N_{11} & N_{12} & 0 & N_{13} & 0 & 0 & 0 \\
            0 & N_{21} & N_{22} & 0 & N_{23} & 0 & 0 & 0 \\
            0 & 0 & 0 & M_{11} & 0 & M_{12} & M_{13} & 0 \\
            0 & N_{31} & N_{32} & 0 & N_{33} & 0 & 0 & 0 \\
            0 & 0 & 0 & M_{21} & 0 & M_{22} & M_{23} & 0 \\
            0 & 0 & 0 & M_{31} & 0 & M_{32} & M_{33} & 0 \\
            0 & 0 & 0 & 0 & 0 & 0 & 0 & e^{\frac{3}{2}i\omega_0 t} \\
          \end{array}
        \right)}_{W(t)}\left(
                 \begin{array}{c}
    C_{+++}(0) \\
    C_{++-}(0) \\
    C_{+-+}(0) \\
    C_{+--}(0) \\
    C_{-++}(0) \\
    C_{-+-}(0) \\
    C_{--+}(0) \\
    C_{---}(0) \\
  \end{array}
               \right),
\end{equation}
where $W(t)$ is the evolution matrix, and $M_{ij}$, $N_{ij}$ are components of the matrices $M$ and $N$, respectively. Let the initial state of the system be a product state
\begin{equation}\label{TLSIni}
 \hat{\rho} (0)=\hat{\rho} _1 (0)\otimes\hat{\rho}^S (0)\otimes\hat{\rho} _2 (0),
\end{equation}
where the main system is initially prepared in an arbitrary state
\begin{equation}\label{initialsys}
 \hat{\rho}^S (0)=\left(
                    \begin{array}{cc}
                      a & \bar{c} \\
                      c & b \\
                    \end{array}
                  \right),\,\,(a+b=1,\,\,(a-b)^2+4c\bar{c}\leq 1),
\end{equation}
and the reservoir states are thermal states given by
\begin{equation}\label{b16}
  \begin{split}
     \hat{\rho} _1 (0)=& \begin{pmatrix}
                     p_{1} & 0 \\
                     0 & q_1
                   \end{pmatrix},\,\,\,(p_1+q_1=1), \\
      \hat{\rho} _2 (0)=& \begin{pmatrix}
                     p_{2} & 0 \\
                     0 & q_2
                   \end{pmatrix},\,\,\,(p_2+q_2=1), \\
  \end{split}
\end{equation}
respectively.
In the standard basis, we have
\begin{equation}\label{totalroxero}
  \hat{\rho} (0)=\left(
                   \begin{array}{cccccccc}
                     p_1p_2 a & 0 & p_1 p_2 \bar{c} & 0 & 0 & 0 & 0 & 0 \\
                     0 & p_1 q_2 a & 0 & p_1 q_2 \bar{c} & 0 & 0 & 0 & 0 \\
                     p_1 p_2 c & 0 & p_1 p_2 b & 0 & 0 & 0 & 0 & 0 \\
                     0 & p_1 q_2 c & 0 & p_1 q_2 b & 0 & 0 & 0 & 0 \\
                     0 & 0 & 0 & 0 & q_1 p_2 a & 0 & q_1 p_2 \bar{c} & 0 \\
                     0 & 0 & 0 & 0 & 0 & q_1 q_2 a & 0 & q_1 q_2 \bar{c} \\
                     0 & 0 & 0 & 0 & q_1 p_2 c & 0 & q_1 p_2 b & 0 \\
                     0 & 0 & 0 & 0 & 0 & q_1 q_2 c & 0 & q_1 q_2 b \\
                   \end{array}
                 \right).
\end{equation}
Therefore, the evolved density matrix is
\begin{equation}\label{totalroht}
  \hat{\rho} (t)=W(t)\hat{\rho} (0) W^\dag(t).
\end{equation}
The reduced density matrix of the oscillator can be obtained by tracing out the bath degrees of freedom
\begin{equation}\label{reduced}
  \hat{\rho}^S (t)=tr_{\text{Baths}}[\hat{\rho} (t)],
\end{equation}
or equivalently in terms of the components
\begin{equation}\label{reducedcomponents}
  \rho^S_{ij} (t)=\sum_{i_1,i_2\in \{+,-\}}\la i_1|\la i|\la i_2|\hat{\rho}(t)|i_1\ra|j\ra|i_2\ra,\,\,\,\,\,(i,j\in\{+,-\}).
\end{equation}
After straightforward calculations, the time evolution of the independent components of the reduced density matrix are obtained as
\begin{eqnarray}\label{Explicitcomponents}
  \rho^S_{++} (t) &=& a\cos^2[\tilde{G}]+\frac{\gamma_1 p_1+\gamma_2 p_2}{\gamma_1+\gamma_2}\,\sin^2[\tilde{G}],\nn\\
  \rho^S_{+-} (t) &=& \frac{\bar{c} e^{-i\omega_0 t}\Big[(\gamma_1^2+\gamma_2^2+2\gamma_1 \gamma_2(p_2 p_1+q_2 q_1))\cos^2[\tilde{G}]+2\gamma_1 \gamma_2 (p_2 q_1+p_1 q_2)\Big]}{(\gamma_1+\gamma_2)^2}.
\end{eqnarray}
In the long-time regime, using $\cos^2[\tilde{G}]=e^{-(\gamma_1+\gamma_2)t}$, these components tend to the stationary values
\begin{eqnarray}\label{MarkovEvo}
  \rho^S_{++} (\infty) &=& \frac{p_1\gamma_1+p_2\gamma_2}{\gamma_1+\gamma_2},\nn\\
   |\rho^S_{+-} (\infty)| &=& \frac{2c\,\gamma_1 \gamma_2 (p_2 q_1+p_1 q_2)}{(\gamma_1+\gamma_2)^2}.
\end{eqnarray}
Therefore, the main two-level system has been thermalized with an equivalent thermal bath with average up state population $\bar{p}=\frac{p_1\gamma_1+p_2\gamma_2}{\gamma_1+\gamma_2}$. Also, in the long time regime, a non zero coherency $|\rho^S_{+-} (\infty)|$ has been induced which is zero in the presence of a single bath $(\gamma_1=0$ or $\gamma_2=0)$.
\subsection{Markovian and non-Markovian process}
A process is said to be non-Markovian if there exists a pair of initial states $\rho_1 (0)$ and $\rho_2 (0)$ and a certain
time $t$ such that \cite{f4,Gardiner2004,Rivas,Laine2014,Budini2018,Vega2017,Tamadcelli2018}
\begin{equation}\label{D1}
 \sigma(t)=\frac{d}{d t} D(\rho_1 (t), \rho_2 (t))>0.
\end{equation}
The trace distance $D(.,.)$ between the evolved states is defined by
\begin{equation}\label{D2}
   D(\rho_1 (t), \rho_2 (t))=\frac{1}{2} tr(|\rho_1 (t)-\rho_2 (t)|),
\end{equation}
where $|\hat{O}|=\sqrt{\hat{O}^\dag \hat{O}}$. For a Hermitian matrix $(\hat{O}^\dag=\hat{O})$, the trace $tr |\hat{O}|$ can be written as a sum over the
absolute values of the eigenvalues of $\hat{O}$, $tr |\hat{O}|=\sum\limits_{i}|\lambda_i|$.

To investigate if the evolution given in Eq.~(\ref{Explicitcomponents}) is Markovian or not, let us find the rate $\sigma(t)$ for the initial states
\begin{equation}\label{sig1}
  \rho_1 (0)=\left(
               \begin{array}{cc}
                 a_1 & 0 \\
                 0 & b_1 \\
               \end{array}
             \right),\,\,\,\,  \rho_2 (0)=\left(
               \begin{array}{cc}
                 a_2 & 0 \\
                 0 & b_2 \\
               \end{array}
             \right),
\end{equation}
the evolved states are
\begin{eqnarray}\label{sig2}
&&  \rho_1 (t)=\left(
               \begin{array}{cc}
                 a_1 \cos^2[\tilde{G}]+\bar{p}\sin^2[\tilde{G}] & 0 \\
                 0 & 1-[a_1 \cos^2[\tilde{G}]+\bar{p}\sin^2[\bar{G}]] \\
               \end{array}
             \right),\nn\\
&&  \rho_2 (t)=\left(
               \begin{array}{cc}
                 a_2 \cos^2[\tilde{G}]+\bar{p}\sin^2[\tilde{G}] & 0 \\
                 0 & 1-[a_2 \cos^2[\tilde{G}]+\bar{p}\sin^2[\tilde{G}]] \\
               \end{array}
             \right),
\end{eqnarray}
and
\begin{equation}\label{diff}
  \rho_1 (t)-\rho_2 (t)=\left(
                          \begin{array}{cc}
                            (a_1-a_2)\cos^2[\tilde{G}] & 0 \\
                            0 & (a_2-a_1)\cos^2[\tilde{G}] \\
                          \end{array}.
                        \right)
\end{equation}
The trace distance is
\begin{equation}\label{D3}
 D(\rho_1 (t), \rho_2 (t))=|a_2-a_1|\cos^2[\tilde{G}].
\end{equation}
and for the distance rate we obtain
\begin{equation}\label{rate2}
  \sigma(t)=\frac{d}{d t} D(\rho_1 (t), \rho_2 (t))=-|a_2-a_1|\sin[2\tilde{G}]\,g(t).
\end{equation}
Therefore, Markovianity or Non-Markovianity depends on the time-dependent coupling function $g(t)$. For the choice $\cos^2[\tilde{G}]=e^{-(\gamma_1+\gamma_2)t}$, the trace distance $D(t)=|a_2-a_1|\,e^{-(\gamma_1+\gamma_2) t}$ is a decreasing function in time ($\sigma(t)<0$), leading to a Markovian process. For a constant coupling function $g(t)=g_0$, we have $\tilde{G}(t)=g_0\sqrt{\gamma_1+\gamma_2}\,t$, and $\sigma(t)=-g_0 |a_2-a_1|\sin[2g_0\sqrt{\gamma_1+\gamma_2}\,t]$, so there are intervals of time where the process is non-Markovian.
\subsubsection{Example 1}
Let the two-level system be initially prepared in its ground state
\begin{equation}\label{Ex1}
  \hat{\rho}^S (0)=\left(
                     \begin{array}{cc}
                       0 & 0 \\
                       0 & 1 \\
                     \end{array}
                   \right),\,\,(a=c=\bar{c}=0, b=1),
\end{equation}
then by using Eqs.~(\ref{totalroxero},\ref{totalroht},\ref{Ex1}), the probability of finding the system in its excited is
\begin{eqnarray}\label{Ex2}
 \rho^S_{++} (t) &=& \frac{p_1\,\gamma_1+p_2\,\gamma_2}{\gamma_1+\gamma_2}\,\sin^2[\tilde{G}(t)],\nn\\
                 &=& \frac{p_1\,\gamma_1+p_2\,\gamma_2}{\gamma_1+\gamma_2}\,(1-e^{-(\gamma_1+\gamma_2)t}).
\end{eqnarray}
For an initially excited state
\begin{equation}\label{Ex3}
  \hat{\rho}^S (0)=\left(
                     \begin{array}{cc}
                       1 & 0 \\
                       0 & 0 \\
                     \end{array}
                   \right),\,\,(a=1, b=c=\bar{c}=0),
\end{equation}
we obtain
\begin{equation}\label{Ex4}
 \rho^S_{++} (t) = e^{-(\gamma_1+\gamma_2) t}+\frac{p_1\,\gamma_1+p_2\,\gamma_2}{\gamma_1+\gamma_2}\,(1- e^{-(\gamma_1+\gamma_2) t}).
\end{equation}
\subsubsection{Example 2}
The time evolution of pure states of the system described by the Hamiltonian Eq.~(\ref{b1}) can also be obtained directly by making use of Eq.~(\ref{evolution}). As an example, let the initial state of the total system be the separable state
\begin{equation}\label{Ex5}
  |\psi(0)\ra=(\sqrt{p_1}\,|+\ra_1+\sqrt{q_1}\,|-\ra_1)\otimes |-\ra\otimes |-\ra_2,\,\,\,(p_1+q_1=1),
\end{equation}
or equivalently
\begin{equation}\label{Ex6}
  \rho(0)=|\psi(0)\ra\la \psi(0)|=\left(
            \begin{array}{cc}
              p_1 & \sqrt{p_1 q_1} \\
              \sqrt{p_1 q_1} & q_1 \\
            \end{array}
          \right)
\otimes \underbrace{\left(
            \begin{array}{cc}
              0 & 0 \\
              0 & 1 \\
            \end{array}
          \right)}_{\rho^S (0)}\otimes \left(
            \begin{array}{cc}
              0 & 0 \\
              0 & 1 \\
            \end{array}
          \right),
\end{equation}
where the left TLS has a coherency $\sqrt{p_1 q_1}$ and the right TLS is prepared in its ground state. By inserting Eq.~(\ref{Ex6}) into Eq.~(\ref{evolution}), we find
\begin{equation}\label{Ex7}
  |\psi(t)\ra=M_{11}\sqrt{p_1}|+\ra|-\ra|-\ra+M_{21}\sqrt{p_1}|-\ra|+\ra|-\ra+M_{31}\sqrt{p_1}|-\ra|-\ra|+\ra+e^{3i\omega_0 t/2}\sqrt{q_1}|-\ra|-\ra|-\ra.
\end{equation}
The reduced density matrix of the main TLS is
\begin{equation}\label{Ex8}
  \rho^S (t)=tr_{1,2}\big(|\psi(t)\ra\la \psi(t)|\big)=\left(
                                                         \begin{array}{cc}
                                                           |M_{12}|^2 p_1 & M_{21}\sqrt{p_1 q_1}e^{-3i\omega_0 t/2} \\
                                                           \bar{M}_{21}\sqrt{p_1 q_1}e^{3i\omega_0 t/2} & 1-|M_{12}|^2 p_1 \\
                                                         \end{array}
                                                       \right),
\end{equation}
where $tr_{1,2}$ denotes taking trace over the degrees of freedom of the left and right TLS. For the choice $\cos^2[G]=e^{-(\gamma_1+\gamma_2)t}$, we have
\begin{eqnarray}\label{Ex9}
  \rho^S_{++} (t) &=& |M_{21}|^2 p_1=\frac{p_1 \gamma_1}{\gamma_1+\gamma_2}\,(1-e^{-(\gamma_1+\gamma_2)t}),\nn \\
      |\rho^S_{+-}(t)| &=& |M_{21}| \sqrt{p_1 q_1}=\Big[\frac{p_1 q_1 \gamma_1}{\gamma_1+\gamma_2}\,(1-e^{-(\gamma_1+\gamma_2)t})\Big]^{\frac{1}{2}}.
\end{eqnarray}
Therefore, in the long-time regime, the main TLS is thermalized with an induced coherency $\sqrt{\frac{p_1 q_1 \gamma_1}{\gamma_1+\gamma_2}}$.
\section{Summary and conclusion}\label{conclusion}
We explored the quantum dynamics of two fundamental systems-a harmonic oscillator and a two-level system-interacting with multiple thermal baths characterized by different temperatures and coupling strengths.
We provided a rigorous analytical treatment of these interactions, revealing key insights into the effects of dissipation and decoherence in such open quantum systems.
For a dissipative harmonic oscillator, we derived an exact analytical solution for its reduced density matrix and utilized characteristic functions to compute essential quantum distributions,
such as the Husimi Q-function, Glauber-Sudarshan P-function, and Wigner function. These results offer a detailed characterization of the oscillator's quantum state in the presence of multiple thermal reservoirs.
Notably, we demonstrated that a system coupled to multiple baths can be effectively mapped onto an equivalent system with a single thermal bath and an appropriately defined effective coupling strength,
simplifying its analytical treatment. Extending our analysis to a two-level system coupled to two thermal reservoirs, we derived the exact time evolution operator and obtained the reduced density matrix,
shedding light on the system's quantum dynamics.
\vspace{20pt}

\section*{Acknowledgements} This work has been supported by the University of Kurdistan. The authors thank Vice Chancellorship of Research and Technology, University of Kurdistan.

\section*{Data availability} This article has no associated data.

\section*{Conflict of Interest} The authors declare no conflicts of interest.


\begin{thebibliography}{99}

\bibitem{a1} Rieder Z, Lebowitz J and Lieb E 1967 {\it Journal of Mathematical Physics} \textbf{8} 1073
\bibitem{a2} Martinez E A and Paz J P 2013 {\it Physical review letters} \textbf{110} 130406
\bibitem{a3} Dhar A 2008 {\it Advances in Physics} \textbf{57} 457
\bibitem{a4} Landi G T and de Oliveira M J 2014 {\it Physical Review E} \textbf{89} 022105
\bibitem{a5} Asadian A, Manzano D, Tiersch M and Briegel H 2013 {\it Physical Review E} \textbf{87} 012109
\bibitem{a6} Fogedby H C and Imparato A 2014 {\it Journal of Statistical Mechanics: Theory and Experiment} \textbf{2012} P04005
\bibitem{a7} Cahill D G, Braun P V, Chen G, Clarke D R, Fan S, Goodson K E, Keblinski P, King W P, Mahan G D,
             Majumdar A, et al. 2014 {\it Applied physics reviews} \textbf{1}
\bibitem{a8} Galve F, Giorgi G L, and Zambrini R 2010 {\it Physical Review A} \textbf{81} 062117
\bibitem{a10} Dhar A 2008 {\it Advances in Physics} \textbf{57} 457
\bibitem{a11} Wu W and An J -H 2024 {\it Physical Review Letters} \textbf{133} 050401
\bibitem{a12} Delle Site L and Hartmann C 2024 {\it Molecular Physics} e2391484
\bibitem{a13} Ghesqui\`{e}re A, Sinayskiy I and Petruccione F 2013 {\it Physics Letters A} \textbf{377} 1682
\bibitem{a15} Akutsu N 2024 {\it Journal of Crystal Growth} \textbf{631} 127610
\bibitem{f1} Esposito M 2012 {\it Physical Review E} \textbf{85} 041125
\bibitem{f2} Spohn H and Lebowitz J L 2007 {\it Adv. Chem. Phys} \textbf{38} 109
\bibitem{f3} Cuetara G B, Engel A, and Esposito M 2002 {\it New journal of physics} \textbf{17} 055002
\bibitem{f4} Breuer H -P and Petruccione F 2002 {\it The theory of open quantum systems} (Oxford University Press).
\bibitem{f5} Kosloff R 2013 {\it Entropy} \textbf{15} 2100
\bibitem{f6} Esposito M, Ochoa M A and Galperin M 2015 {\it Physical Review B} \textbf{92} 235440
\bibitem{x1} Schlosshauer M, Hines A P and Milburn G J 2008 {\it Physical Review A} \textbf{77} 022111
\bibitem{x2} Militello B Nakazato H and Napoli A 2017 {\it Physical Review A} \textbf{96} 023862
\bibitem{x3} Maimbourg T 2024 {\it Physical Review B} \textbf{110} 064203
\bibitem{x4} Kheirandish F, Cheraghpour N, and Moradian A 2024 arXiv:2412.03943
\bibitem{c1} Bhattacharya J, Gangopadhyay G and Gangopadhyay S 2025 {\it Physica Scripta} \textbf{100} 025103
\bibitem{c2} Weinbub J and Kosik R 2022 {\it Journal of Physics: Condensed Matter} \textbf{34} 163001
\bibitem{c3} Shaker L M, Al-Amiery A, Isahak W N R W and Al-Azzawi W K 2023 {\it Journal of Optics} \textbf{1}
\bibitem{b1} St\"{o}ber J, B\"{a}cker A and Ketzmerick R 2024 {\it Physical Review Letters} \textbf{132} 047201
\bibitem{b2} Fedorova A V and Yurischev M A 2021 {\it Quantum Information Processing} \textbf{20} 169
\bibitem{b3} Nettersheim J, Burgardt S, Bouton Q, Adam D, Lutz E and Widera A 2022 {\it PRX Quantum} \textbf{3} 040334
\bibitem{b4} Alsulami M and Abd-Rabbou M 2024 {\it Annalen der Physik} \textbf{536} 2400122
\bibitem{d22} Kheirandish F, Bolandhemmat E, Cheraghpour N, Moradi R and Ahmadian S 2024 {\it Physica Scripta} \textbf{100} 015110
\bibitem{gelbwaser2013minimal} Gelbwaser-Klimovsky D, Alicki R and Kurizki G 2013 {\it Physical Review E} \textbf{87} 012140
\bibitem{Louisell} Louisell W H (1973) {\it Quantum statistical properties of radiation} (John Wiley and Sons, Inc.)
\bibitem{GN} Gerry C C and Knight P L 2023 {\it Introductory quantum optics} (Cambridge university press)
\bibitem{ban2019decoherence} Ban M 2019 {\it Quantum Information Processing} \textbf{18} 220
\bibitem{ram2022} Wolfram Research 2022 {\it Mathematica} \textbf{13.1}
\bibitem{Gardiner2004} Gardiner C W and Zoller P 2004 Quantum noise: A handbook of Markovian and Non-Markovian Quantum Stochastic Methods with
Applications to Quantum Optics (3rd ed) (Springer Science and Business Media)
\bibitem{Rivas} Rivas A, Huelga S F and Plenio M B 2014 Quantum non-Markovianity: characterization, quantification and detection Rep. Prog. Phys.
77 094001
\bibitem{Laine2014} Laine E M, Piilo J and Breuer H P 2014 Measure for the degree of non-Markovian behavior of quantum processes in open systems Phys.
Rev. Lett. 108 210402
\bibitem{Budini2018} Budini A A 2018 Maximally non-Markovian quantum dynamics without environment-to-system backflow of information Phys. Rev. A
97 052133
\bibitem{Vega2017} De Vega I and Alonso D 2017 Dynamics of non-Markovian open quantum systems Rev. Mod. Phys. 89 015001
\bibitem{Tamadcelli2018} Tamascelli D, Smirne A, Huelga S F and Plenio M B 2018 Nonperturbative treatment of non-Markovian dynamics of open quantum
systems Phys. Rev. Lett. 120 030402

\end{thebibliography}
\end{document}